# Intraoral Pressure in Ethnic Wind Instruments

## Clinton F. Goss

Westport, CT, USA. Email: clint@goss.com

### ABSTRACT

High intraoral pressure generated when playing some wind instruments has been linked to a variety of health issues. Prior research has focused on Western classical instruments, but no work has been published on ethnic wind instruments. This study measured intraoral pressure when playing six classes of ethnic wind instruments ($N = 149$): Native American flutes ($n = 71$) and smaller samples of ethnic duct flutes, reed instruments, reedpipes, overtone whistles, and overtone flutes. Results are presented in the context of a survey of prior studies, providing a composite view of the intraoral pressure requirements of a broad range of wind instruments. Mean intraoral pressure was 8.37 mBar across all ethnic wind instruments and $5.21 \pm 2.16$ mBar for Native American flutes. The range of pressure in Native American flutes closely matches pressure reported in other studies for normal speech, and the maximum intraoral pressure, 20.55 mBar, is below the highest subglottal pressure reported in other studies during singing. Results show that ethnic wind instruments, with the exception of ethnic reed instruments, have generally lower intraoral pressure requirements than Western classical wind instruments. This implies a lower risk of the health issues related to high intraoral pressure.

## Introduction

Intraoral pressure is an important physiological metric related to playing wind instruments. The range of intraoral pressure generated when playing an instrument is dependent on the instrument. Within that range, players alter their breath pressure to control the volume, pitch, and tone produced by the instrument.

Intraoral pressure is a consideration when choosing a wind instrument to play. Wind instruments with high intraoral pressure requirements have been linked to a number of health issues, in particular:

- Velopharyngeal incompetency (VPI), a condition where the soft palate or pharyngeal walls fail to separate the nasal cavity from the oral cavity ([Weber-J 1970], [Dibbell 1979], [Dalston 1988], [Ingrams 2000], [Schwab 2004], [Stasney 2003], [Malick 2007], [Kreuter 2008], [Evans-A 2009], [Evans-A 2010], [Evans-A 2011]).

- Pneumoparotid, where the parotid gland becomes enlarged due to air insufflation ([Kirsch 1999], [Kreuter 2008], [Lee-GG 2012]).

- Hemoptysis ([Kreuter 2008]).

- Increased intraocular pressure and intermittent high-pressure glaucoma ([Schuman 2000], [Schmidtmann 2011]).

- Hypertension (possibly — see [Dimsdale 1995] and [Larger 1996]).

- Barotrauma causing reduced pulmonary function ([Deniz 2006]).

- Laryngocele, a congenital lung condition seen in glassblowers due to high intraoral pressure ([Kreuter 2008]). [Lee-GG 2012] reports intraoral pressure as high as 200 mBar for glassblowing.

Given the range and severity of these potential health issues, preference for wind instruments with lower intraoral pressure requirements is prudent.

Intraoral pressure has been studied in speech, singing, and playing various wind instruments of the Western classical tradition. However, no studies measuring intraoral pressure in ethnic wind instruments have been reported in the literature.

This study was undertaken to determine the intraoral pressure involved in playing a wide range of ethnic wind instruments. Measurements were made on 149 ethnic wind instruments in situations that approximate normal as well as extreme playing techniques. The results are combined with intraoral and subglottal pressure measurements of speech, singing, and other instruments from prior studies in a set of charts that provide a composite view of the intraoral pressure requirements for a broad range of wind instruments. Data tables are also provided for all pressure measurements from this study as well as prior studies.

## Method

### Musical Instruments

Six classes of instruments were studied[1]:

**Native American flutes** (*n* = 71): A front-held flute that has an external block and an internal wall that separates an air chamber from a resonating chamber that contain open finger holes ([Goss 2011]). Hornbostel–Sachs (HS) class 421.211.12 and 421.221.12 — edge-blown aerophones, with breath directed through an external or internal duct against an edge, with finger holes. Native American flutes used in this study were crafted by 32 different flute makers. These flutes play primarily in the first register, with some flutes having a few notes in the second register. Range is typically limited to 12–15 semitones.

**Ethnic duct flutes** (*n* = 46). HS class 421.221.12 — edge-blown aerophones, with breath directed through an external or internal duct against an edge, with finger holes. Typical play on these instruments is done in the first register, with normal play extending to several notes in the second register and possibly one note in the third register (sounding a major twelfth). Ethnic duct flutes in this study include the Irish whistle, Slovakian pistalka (píšťalka), Ukrainian and Russian sopilkas (Сопілка, Сопел), Romanian frula, Indonesian suling, Georgian salamuri (სალამური), Bolivian tarka, Mesoamerican clay flutes, flutes characteristic of the Tarahumara culture, and Russian sivril.

**Ethnic reed instruments** (*n* = 4). HS class 422.1 and 422.2 — reed aerophones, with breath directed against one or two lamellae (reeds) which vibrate and set the air in a resonating chamber in motion. They are limited to play in one register and typically have a limited range of no more than 14 semitones. Ethnic reed instruments measured in this study comprise a Russian jaleika, an Armenian duduk (Դուդուկ), a Kenyan bungo'o, and a bamboo saxophone.

**Ethnic reedpipes** (*n* = 12). HS class 422.31 and 422.32 — reed aerophones, single or double reedpipe with a free reed that vibrates through/at a closely fitted frame, with finger holes. They are limited to play in one register and typically have a limited range of no more than 14 semitones. Note that pitch on these instruments often responds inversely to ethnic duct flutes — decreasing as breath pressure is increased. Ethnic reedpipes measured in study comprise the Chinese bawu (巴乌), Chinese hulusi (葫芦絲), and Laotian kaen or khene (ແຄນ).

**Ethnic overtone whistles** (*n* = 8). HS class 421.221.12 — edge-blown aerophones, with breath directed through an internal duct against an edge, without finger holes. Due to the lack of finger holes, they have a fixed-length resonating chamber. They are designed to play high into the overtone series — sometimes as high as the tenth register. To accomplish this, they tend to have relatively long resonating chambers compared with their diameter. Ethnic overtone whistles measured in this study include the Slovakian koncovka, Norwegian willow flute (seljefløyteta), and an overtone flute of the North American Choctaw culture.

**Ethnic overtone flutes** (*n* = 8). HS class 421.221.12. These flutes share some of the characteristics of ethnic duct flutes and ethnic overtone flutes: they have a limited number of finger holes (typically three or four) and are designed to play high into the overtone series. Ethnic overtone flutes measured in this study include the Slovakian fujara, tabor pipe, and flutes of the North American Papago and Pima cultures.

### Measurement

Intraoral pressure was measured using a system constructed of components supplied by Omega Engineering (Stamford, CT). The setup consisted of:

- Meter: one DP25B-S strain meter.
- Sensor: one PX26-001DV piezoresistive pressure sensor, designed for wet conditions. It provides a differential voltage proportional to the pressure applied, in the range of 0–1 PSI.
- One CX136-4 wiring harness.
- Tubing: Clear plastic flexible tubing, 5/16″ outside diameter, 3/16″ inside diameter (55″ for measurements, 128″ for water-column calibration).

---

[1] In some cases, instruments are identified with the culture that initiated the design of the instrument, or the predominant culture where the instrument is presently found. These are provided solely for the purpose of identifying the instrument.

The meter was configured with settings provided by Omega engineering to provide readings in the range $0.001 - 1.000$ PSI in thousandths of a PSI. Based on the combined specifications of the sensor and the meter, the factory calibration of the system should be within $\pm 2.20\%$. This was confirmed by calibrating the unit against the differential height of columns of water in an extended section of tubing, at four pressure points. The greatest deviation was $+2.05\%$.

All readings were converted to milliBars (mBar), including readings from cited sources that are given in a wide variety of units, including cm $H_2O$, in $H_2O$, mm HG, kPa, and psi.

### Procedures

All measurements were taken at 72 °F on instruments that were fully acclimated to that ambient room temperature. Movable parts of an instrument were adjusted to their typical or recommended playing position. Each instrument was warmed up using two long breaths into the finger holes.

The open end of the tubing, cut square, was placed in the mouth perpendicular to the general airflow while the musical instrument was played.

The procedure varied depending on the class of instrument:

**Native American flutes**. Nine measurements were attempted for each flute, three measurements on each of three notes:

- The root note, typically fingered 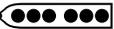

- The fifth note, seven semitones above the root note, typically fingered 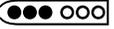 or 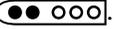. For flutes tuned to the diatonic major scale, the 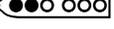 fingering was typically used.

- The octave note, typically fingered 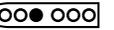 or 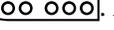. For flutes tuned to the diatonic major scale, the fingering 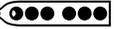 was typically used.

These three notes were played at three dynamics (volumes) by varying breath pressure: *forte* (*f*), *mezzo-forte* (*mf*), and *piano* (*p*). Rather than attempting to produce these dynamics subjectively, a Korg OT-120 pitch meter was used, set to A=440 Hz, equal temperament. A reference pitch (RP) for each note was established based on an "on-pitch" indication on the pitch meter. In the case of instruments that were not tuned to concert pitch, the RP was established using a breath pressure that subjectively produced a good tone.

Pressure readings for *mf* were taken after establishing a steady tone, with no vibrato, that produced the RP. Readings for *f* and *p* were taken with breath pressure that produced readings of 30 cents above and below the RP, respectively. The readings do not show effects of any articulation at the start of the note.

In some cases, it was not possible to produce all nine combinations of pitches and dynamics. For example, increasing breath pressure above *mf* on the root note on some flutes causes the flute to jump into the next register. On some diatonic flutes, readings were not possible when attempting to play *p* on the octave note, since the flute could not maintain resonance in the second register at lower breath pressure.

Repeatability was evaluated by replicating measurements on three flutes on three separate days. The average Coefficient of Variation (CV) was 7.5% with the maximum CV of 11.1%.

**Ethnic duct flutes**. Measurements were taken as with Native American flutes. Most of these flutes use the fingering 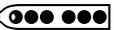 to produce the octave not in the second register. In addition, the fundamental note in each of the higher registers was attempted, as high as was possible on the instrument. Pressure measurements in these higher overtone registers were taken by establishing the pitch and then reducing breath pressure slightly to a point where the tone was stable — reference to precise tuning was not used in these higher overtone registers.

**Ethnic reed instruments**. Measurements were taken as with Native American flutes, except that measurements were taken only for the *mf* dynamic.

**Ethnic reedpipes**. Measurements were taken as with Native American flutes, but it was only possible to use breath pressure to bend pitch $\pm 30$ cents on one instrument. Therefore, most measurements were taken at the *mf* dynamic.

**Ethnic overtone whistles**. One pressure measurement was taken for each note in each register by establishing the pitch and then reducing breath pressure slightly to a point where the tone was stable — reference to precise tuning was not used for this class of instruments.

**Ethnic overtone flutes**. Measurements were taken as with ethnic duct flutes, except that the measurement for the fifth note was taken using the fingering 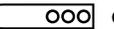 or 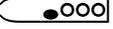 in the first register, regardless of what pitch was produced.

### Literature Survey

A literature search was done for published articles involving intraoral and subglottal measurements in musical instruments, speech, and singing. Pressure measurements from various sources were obtained from numerical data, if published, or by physically interpolating the location of charted data points. This included linear interpolation as well as non-linear interpolation in some cases where graphs used a logarithmic scale. All data points were converted to mBar.

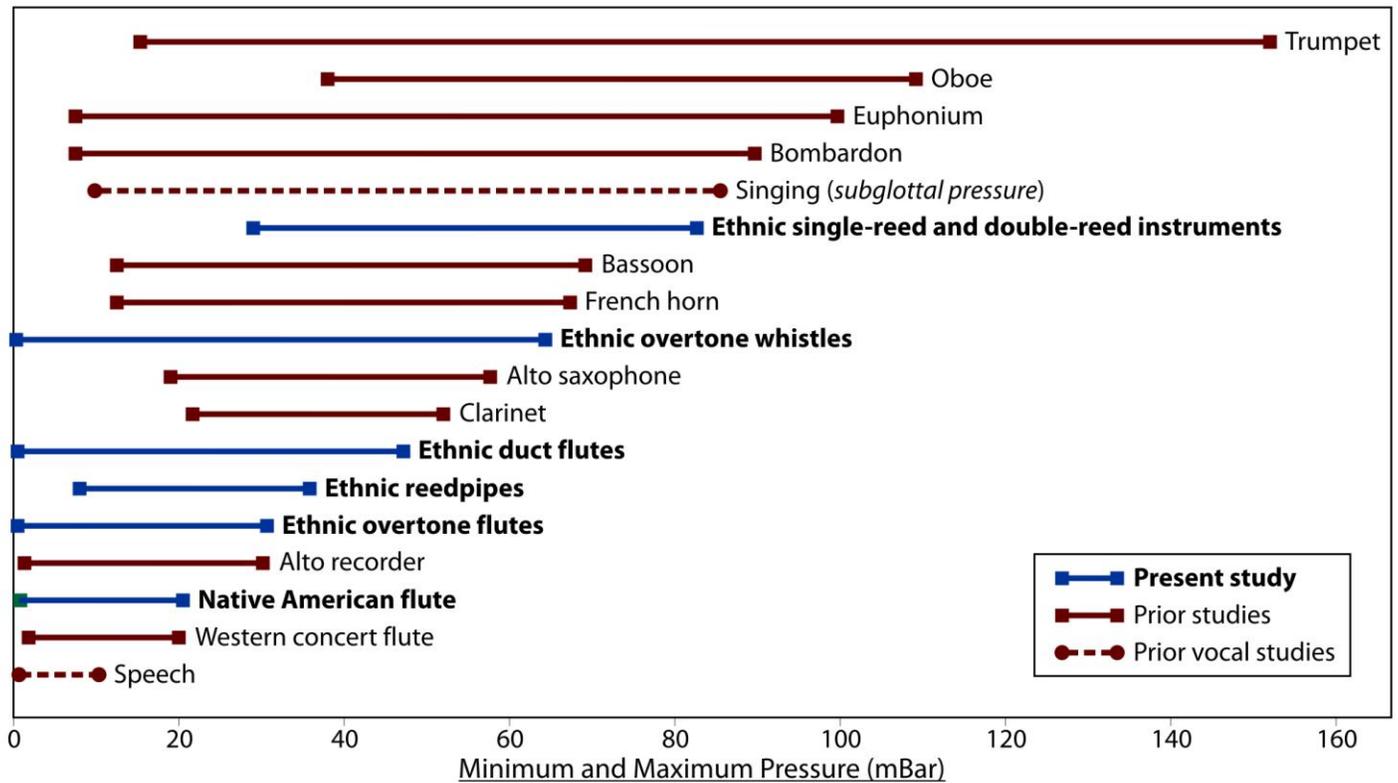

**Figure 1**. Summary of intraoral pressure ranges in wind instruments. The horizontal axis plots the minimum and maximum intraoral pressure for each instrument or instrument class. Measurements from this study are combined with results reported in the literature cited in Tables 1 and 2 contained in the Appendix.

## Results

The composite pressure (CP) for a given instrument is the mean of all measurements for that instrument, including measurements at the various pitches and dynamics that were attainable. The mean intraoral pressure for a group of instruments is the mean of the CP values for the instruments in that group.[2]

The mean intraoral pressure of all ethnic wind instruments in this study was 8.37. Because the CP values across the range of instruments in this study do not show a normal distribution, no standard deviation is reported.[3]

Figure 1 provides an overview of the minimum and maximum intraoral pressure for each instrument or class of instruments, including measurements from this study and a survey of available literature. See Tables 1 and 2 in the Appendix for the source of all data points from prior studies.

Subsequent figures plot pitch on the horizontal axis, grouped into half-octave ranges. For example: $C_3$–$F_3$, $F^\#_3$–$B_3$, ... , $C_6$–$F_6$, and $F^\#_6$–$B_6$ with data points plotted at D and $G^\#$ within each range. The exception is a single measurement at the bottom of the set of half-octave ranges. In that case, the data point is plotted at the actual concert pitch for that measurement.

Intraoral pressure measured on Native American flutes ranged from a minimum of 0.83 mBar to a maximum of 20.55 mBar. The mean intraoral pressure across all Native American flutes was 5.21 ± 2.16 mBar. Because of limitations on some flutes noted previously, of the 639 possible combinations of pitches and dynamics on 71 flutes, 605 actual measurements were taken.

Figure 2 charts the mean intraoral pressure at the **$f$** dynamic (+30 cents) and the **$p$** dynamic (−30 cents), as well as the maximum and minimum of measurements at those dynamics, respectively. See Table 3 in the Appendix for all data values plotted on Figure 2.

Figure 3 places the average **$f$** and **$p$** results for Native American flutes in the context of intraoral and subglottal pressure measurements for speech that have been reported in prior studies.

---

[2] This approach to the analysis was taken – rather than simply averaging all measurements from the group of instruments – since: (a) different instruments contributed different numbers of measurements (because of the limitations of some instruments as noted in the Procedures section) and (b) because the coefficients of variation of measurements for a given flute were reasonably low – averaging 54.3%.

[3] The standard deviation of the CP values for all instruments calculated by traditional methods is ±8.58 mBar.

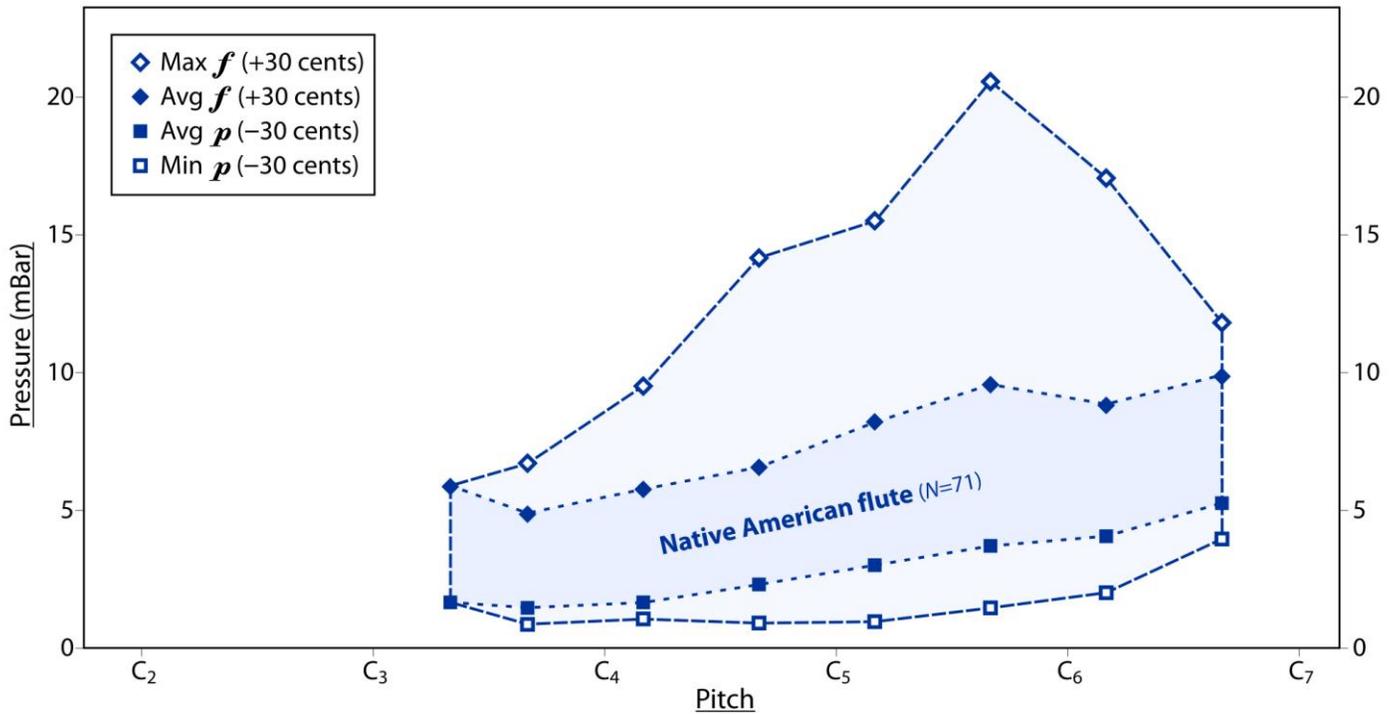

**Figure 2**. Native American flute – intraoral pressure. Pitches are grouped into half-octave ranges, except for the single low measurement taken at $E_3$. $f$ = $forte$, measured at reference pitch (RP) + 30 cents. $p$ = $piano$, measured at RP – 30 cents. RP is concert pitch based on A=440 or, for instruments not tuned to concert pitch, a breath pressure that subjectively produced a good tone.

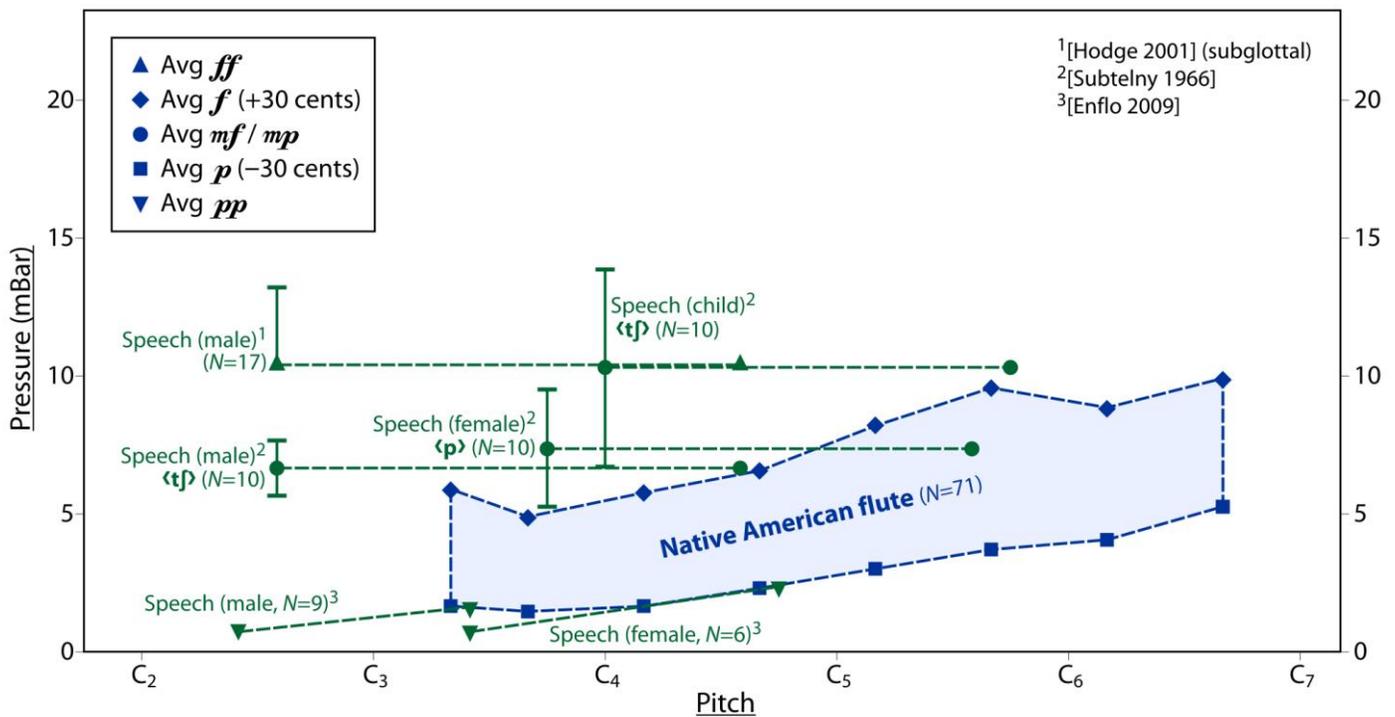

**Figure 3**. Speech and Native American flutes – intraoral pressure. Pitches for Native American flutes are grouped into half-octave ranges, except for the single low measurement taken at $E_3$. $f$ = $forte$, measured at reference pitch (RP) + 30 cents. $p$ = $piano$, measured at RP – 30 cents. RP is concert pitch based on A=440 or, for instruments not tuned to concert pitch, a breath pressure that subjectively produced a good tone. $ff$ ($fortissimo$), $pp$ ($pianissimo$), and $mf$ / $mp$ ($mezzo-forte$ / $mezzo-piano$) are general indications of dynamics that are defined specifically in each of the cited references. The error bars for speech represent one standard deviation.

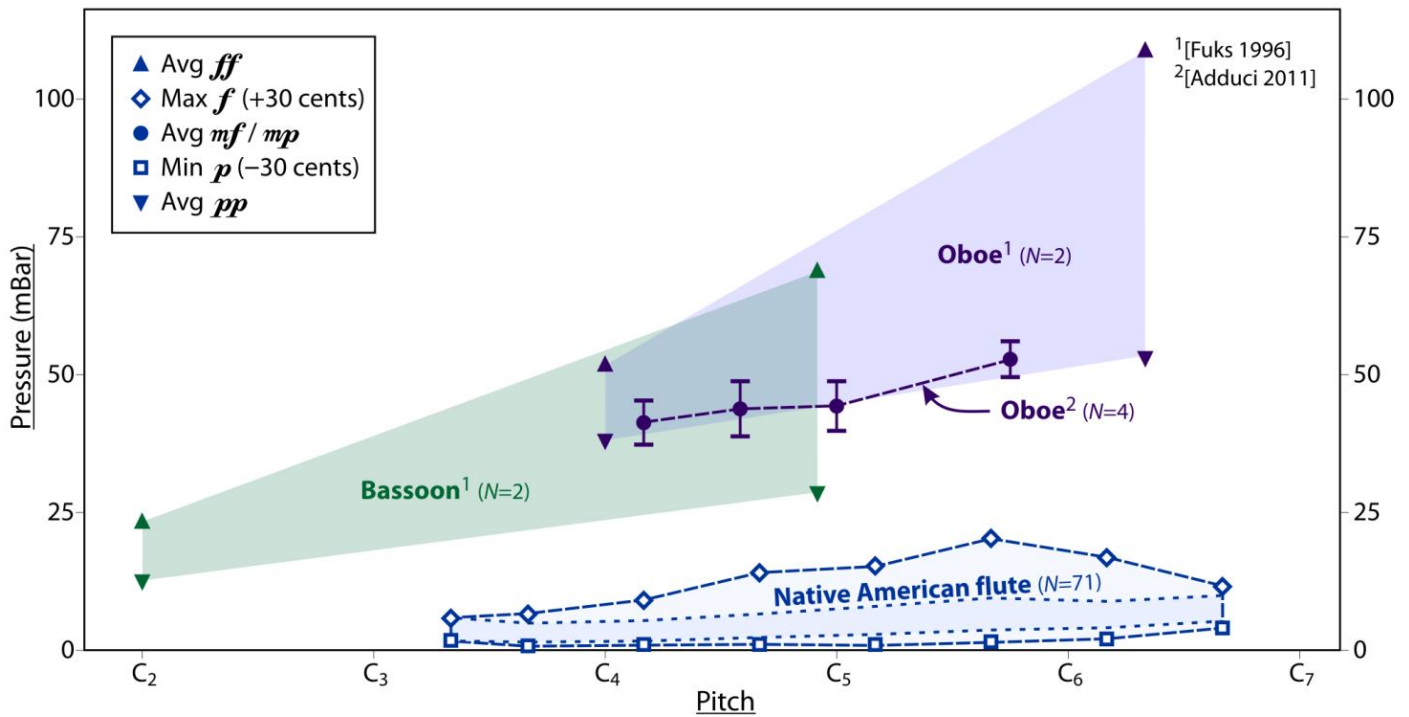

**Figure 4**. Double-reed instruments and Native American flute – intraoral pressure. Pitches for Native American flutes are grouped into half-octave ranges, except for the single low measurement taken at E₃. $f = forte$, measured at reference pitch (RP) + 30 cents. $p = piano$, measured at RP – 30 cents. RP is concert pitch based on A=440 or, for instruments not tuned to concert pitch, a breath pressure that subjectively produced a good tone. $ff$ (*fortissimo*), $pp$ (*pianissimo*), and $mf / mp$ (*mezzo-forte / mezzo-piano*) are general indications of dynamics that are defined specifically in each of the cited references. The error bars represent one standard deviation.

The subglottal measurements reported in [Hodge-FS 2001] for male loud speech (at 75% of their maximum dynamic range) does not have associated pitch information, so Figure 3 uses the typical male pitch range from [Williams-J 2010]. Likewise, the intraoral pressure measurements for consonants from [Subtelny 1966], table 1 (as cited in [Baken 2000], table 8-5) use the suggested pitch ranges for males, females and children from [Williams-J 2010]. [Enflo 2009] provides measurements for the lower limits of phonation at which speech becomes possible. Note that only the lower two frequencies of the male and female lower phonation limits are plotted. See Tables 2 and 3 in the Appendix for all data values plotted on Figure 3.

Figure 4 introduces a change in the vertical scale of pressure by a factor of five to accommodate higher pressure for musical instruments reported in the literature. Pressure measurements are plotted for the bassoon and oboe from two sources.

Figure 5 plots reported measurements on four additional Western classical instruments: the clarinet, alto saxophone, Western concert flute, and the alto recorder. See Table 1 in the Appendix for numeric data values.

Mean intraoral pressure on ethnic duct flutes was 7.26 ± 3.93 mBar and spanned a range of 0.48–47.23 mBar. This range includes what might be considered extreme playing techniques on these instruments, since they were played as high as the ninth register in keeping with the procedures for this study. The subset of measurements on these instruments limited to playing at the reference pitch in registers 1–3, what might be considered a normal range of play on these instruments, gives a mean intraoral pressure of 5.75 ± 3.29 mBar and spanned a range of 0.48–25.86 mBar.

Figure 6 plots the results across the pitch range of ethnic duct flutes. It also plots subglottal pressure measurements from the literature for singing. See Tables 2 and 4 in the Appendix for all data values plotted on Figure 6.

Figure 7 plots the profile for three classes of instruments from this study:

- Ethnic reed instruments had a mean intraoral pressure of 50.38 ± 11.45 mBar and a range of 28.96–82.74 mBar.

- Ethnic reedpipes had a mean intraoral pressure of 18.07 ± 4.26 mBar and a range of 7.93–35.85 mBar.

- Ethnic overtone whistles had a mean intraoral pressure of 11.01 ± 4.84 mBar and a range of 0.28–64.26 mBar.

See Tables 5, 6, and 7 in the Appendix for all data values plotted on Figure 7.

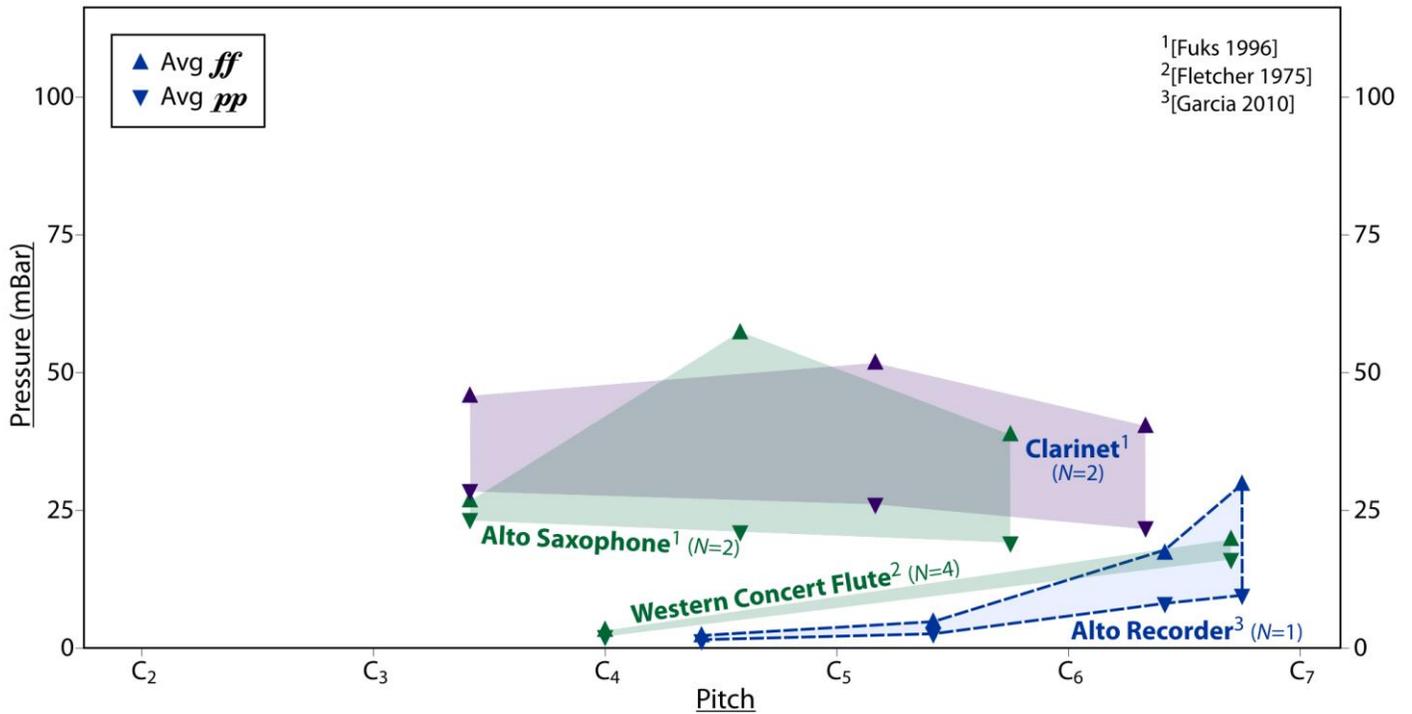

**Figure 5**. Woodwind instruments – intraoral pressure. *ff* (*fortissimo*), *pp* (*pianissimo*), and *mf* / *mp* (*mezzo-forte / mezzo-piano*) are general indications of dynamics that are defined specifically in each of the cited references.

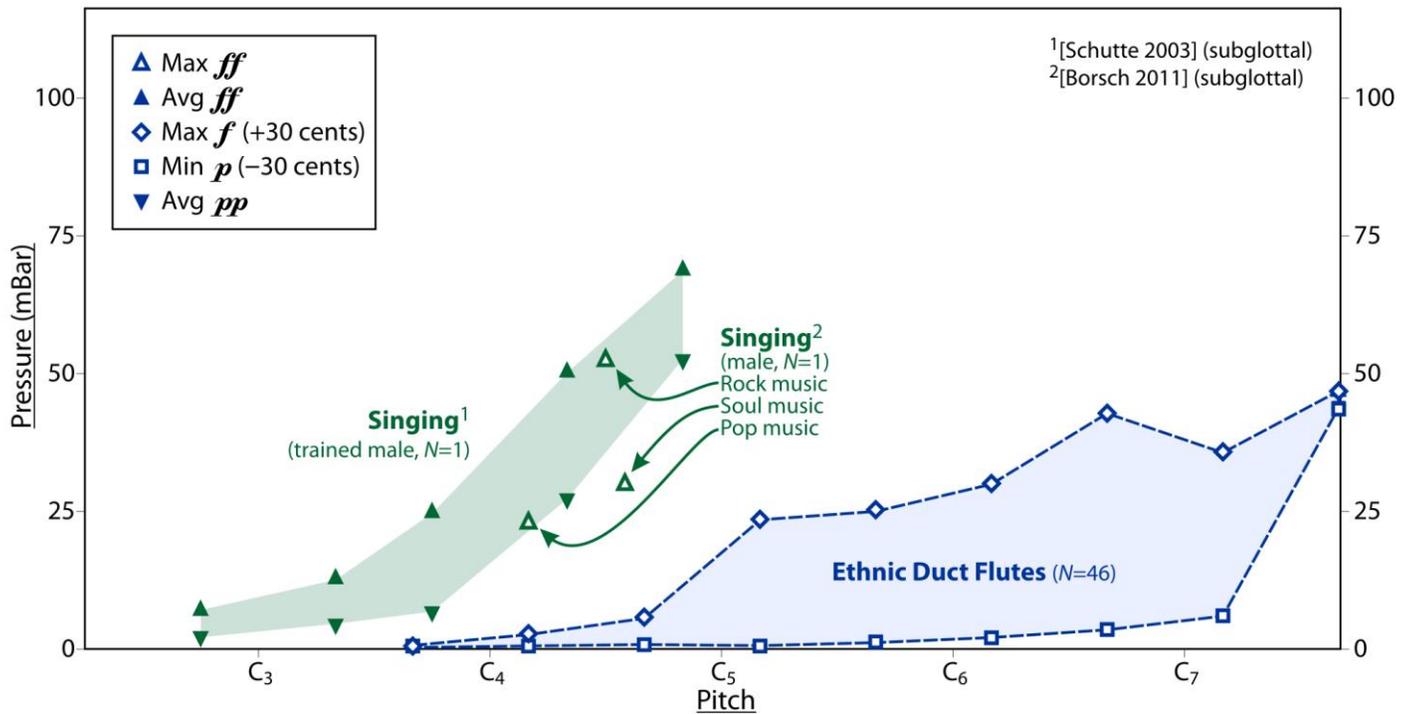

**Figure 6**. Ethnic duct flutes and singing – intraoral and subglottal pressure. Pitches for ethnic duct flutes are grouped into half-octave ranges. *f* = *forte*, measured at reference pitch (RP) + 30 cents. *p* = *piano*, measured at RP – 30 cents. RP is concert pitch based on A=440 or, for instruments not tuned to concert pitch, a breath pressure that subjectively produced a good tone. *ff* (*fortissimo*) and *pp* (*pianissimo*) are general indications of dynamics that are defined specifically in each of the cited references.

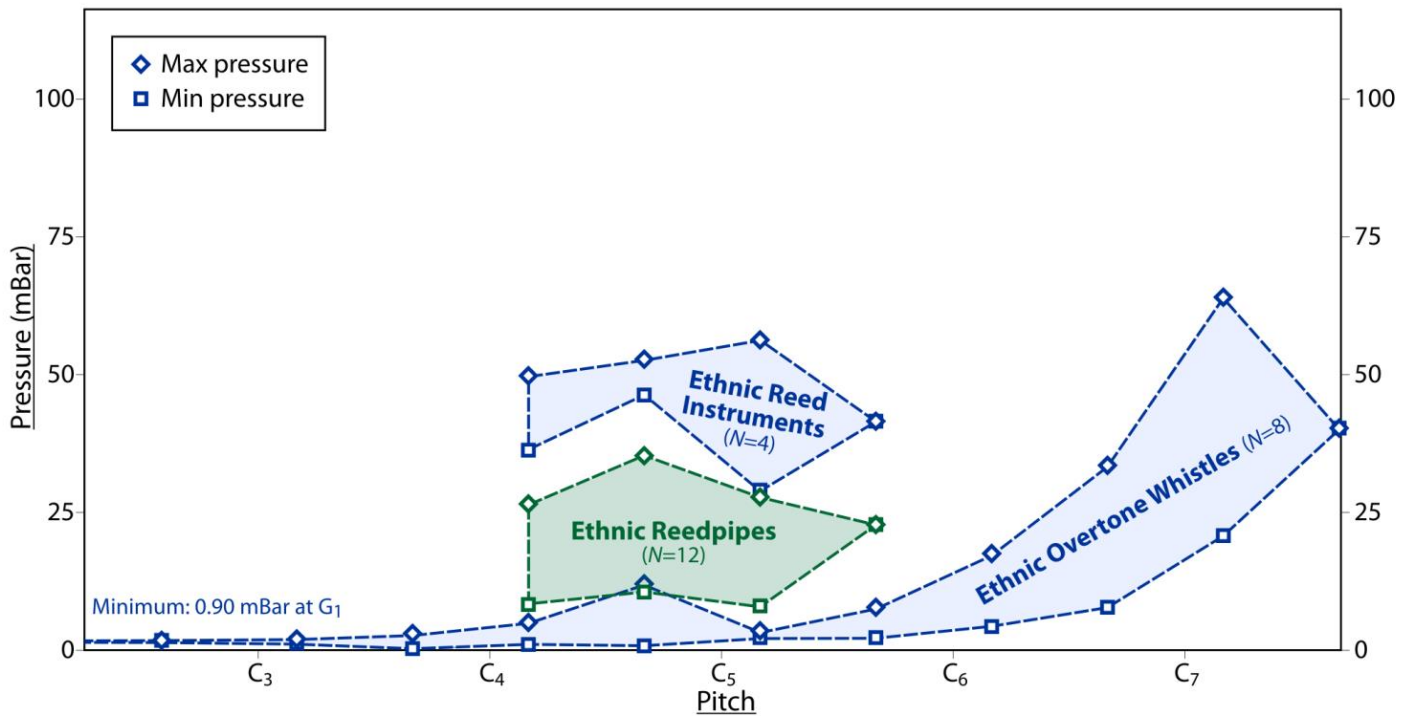

**Figure 7**. Ethnic wind instruments – intraoral pressure. Pitches are grouped into half-octave ranges, with each indicator showing the maximum or minimum intraoral pressure recorded across all measurements taken in that half-octave pitch range.

Mean intraoral pressure on ethnic overtone flutes was $6.94 \pm 3.14$ mBar and spanned a range of 0.55–30.68 mBar. Figure 8 plots the results across the pitch range of these flutes. It also plots intraoral pressure measurements for four brass instruments from the literature. See Table 8 in the Appendix for all data values plotted on Figure 8.

### Breath Pressure Profile

In addition to the primary focus of this study, the measurements collected can shed light on some other issues of wind instrument design. One relates to the concept of a breath pressure profile (BPP) – the graph of intraoral pressure requirements as a player ascends the instruments scale.

Figure 9 charts the BPP for a subgroup of 67 Native American flutes.[4] The lines connect data points for the root, fifth, and octave notes of the same dynamic.

This chart shows that, on the average, Native American flutes are constructed assuming a modest increase in breath pressure as the player moves up the scale, from 3.72 mBar at the root, to 4.74 mBar on the fifth, to 5.40 mBar on the octave note. It also demonstrates that the increase in breath pressure is, on average, linear through the three notes measured and that the linear relationship holds across changes in dynamics.

Figure 9 also shows that a larger change in breath pressure is needed to raise pitch by 30 cents from concert pitch than to lower pitch by 30 cents. These results use the average readings across the three pitches at each of the dynamics: from the average  intraoral pressure of 4.62 mBar for concert pitch, raising pitch by 30 cents required 3.42 mBar more pressure (+74.0%) and lowering pitch by 30 cents required 1.70 mBar less pressure (−36.9%).

Figure 10 shows another BPP plot of the primary notes for a single, well-tuned, six-hole, Native American flute. The lines on this plot are straight and pass through the pressure measurements for the root and the octave notes. The middle line for the *mf* readings at RP shows a slight increase in breath pressure as the player ascends the scale. Given that the upper and lower lines represent a deviation of 30 cents, it is apparent that the variations in tuning on each of the primary notes across the range of the instrument are no more than a few cents.

### Discussion

Figure 2 highlights an interesting comparison between the intraoral pressure involved in speech and playing Native American flutes. This class of flutes grew from a tradition of poetic speech ([Nakai 1996], page 41), and Figure 2 lends empirical evidence to that historical link. Many aspects of speech do not involve intraoral pressure, since the mouth is often open. However, the limits of subglottal pressure and intraoral pressure involved in male loud speech, plosives and fricatives, and the lower phonation thresholds provide a striking correspondence to the limits of intraoral pressure for Native American flutes measured in this study.

---

[4] The four Native American flutes that could not maintain resonance at the *p* dynamic were excluded from this subgroup.

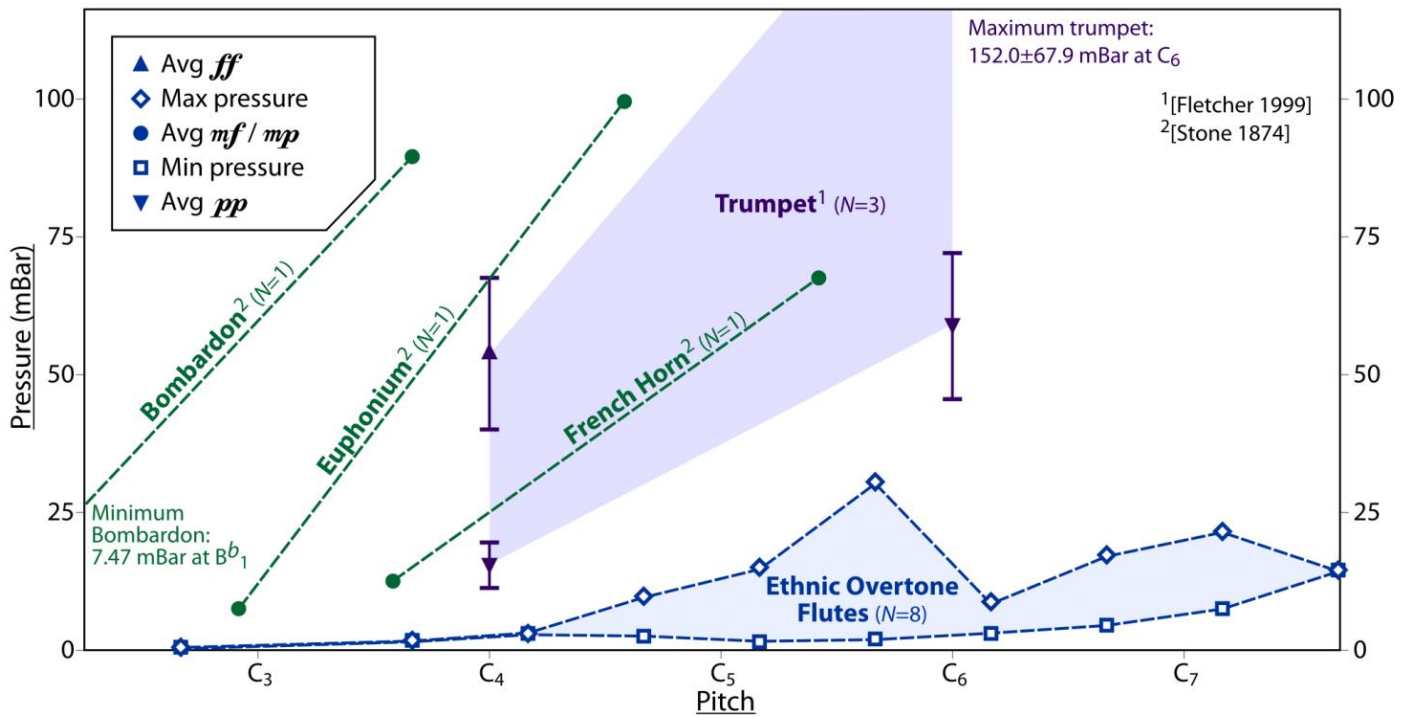

**Figure 8.** Brass instruments and ethnic overtone flutes – intraoral pressure. Pitches for ethnic overtone flutes are grouped into half-octave ranges, with each indicator showing the maximum or minimum intraoral pressure recorded across all measurements taken in that half-octave pitch range. *ff* (*fortissimo*), *mf / mp* (*mezzo-forte / mezzo-piano*), and *pp* (*pianissimo*) are general indications of dynamics that are defined specifically in each of the cited references.

A comparison between Figures 4, 5, and 7 shows that the intraoral pressure involved in playing ethnic reed instruments are roughly aligned with those of Western double-reed and single-reed instruments. Lower intraoral pressure was observed in ethnic reedpipes. It is interesting that the class of ethnic reedpipes includes instruments such as the Chinese bawu and hulusi that are generally widespread in use with amateur rather than professional musicians.

Play on ethnic duct flutes shows a rather large range on Figure 6. However, restricting play to a normal range on these instruments reduces the charted maximum intraoral pressure from 47.23 mBar to 25.86 mBar. This places ethnic duct flute roughly aligned with the one example of a Western duct flute, the alto recorder, charted in Figure 5.

The two classes of ethnic overtone instruments charted in Figures 7 and 8 show that, even though these instruments can perform in a wide range of registers and pitches, they generally have low intraoral pressure requirements. The exception for these instruments is high pressure in ethnic overtone flutes in the extreme upper registers of these instruments. Given that these upper registers are used very briefly in typical play on these instruments, the transient use of these pressure spikes may allow players to avoid the health issues associated with high intraoral pressure.

Aside from the class of ethnic reed instruments that were part of this study, it appears that ethnic wind instruments, in general, have lower intraoral pressure requirements than most Western classical wind instruments.

The development of the concept of breath pressure profile in Figures 9 and 10 demonstrate that an intraoral pressure meter could be a significant aid to tuning wind instruments. The current practice among makers of these instruments is to use their own subjective personal preferences for breath pressure when tuning the instrument across the range of pitches. The use of an intraoral pressure meter would allow the maker to choose a desired breath pressure profile and objectively tune to that specific profile.

### Limitations

Because the meter used in this study required a steady-state pressure of at least 333 msec, short-term intraoral

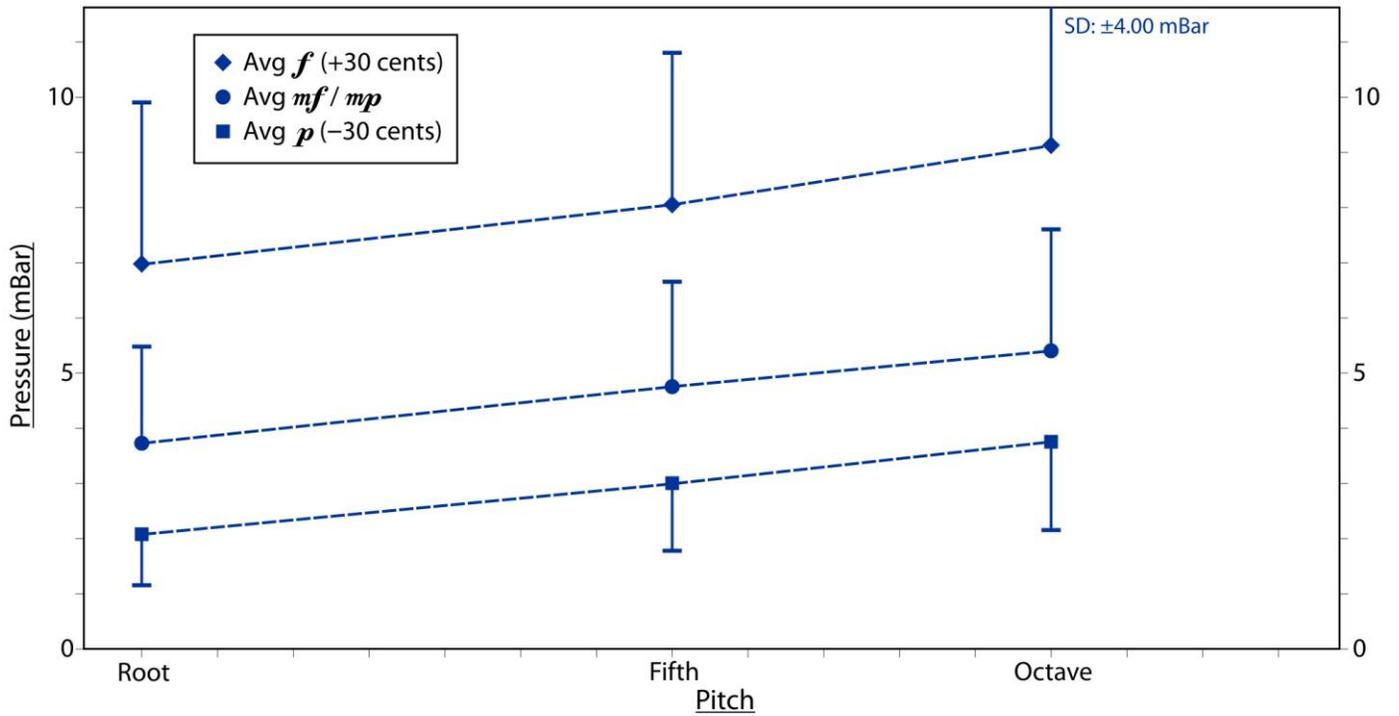

**Figure 9.** Breath pressure profile across the Native American flute range (*N* = 67). *f* = *forte,* measured at reference pitch (RP) + 30 cents. *mf* / *mp* = *mezzo-forte / mezzo-piano,* measured at RP. *p* = *piano,* measured at RP – 30 cents. RP is concert pitch based on A=440 or, for instruments not tuned to concert pitch, a breath pressure that subjectively produced a good tone.

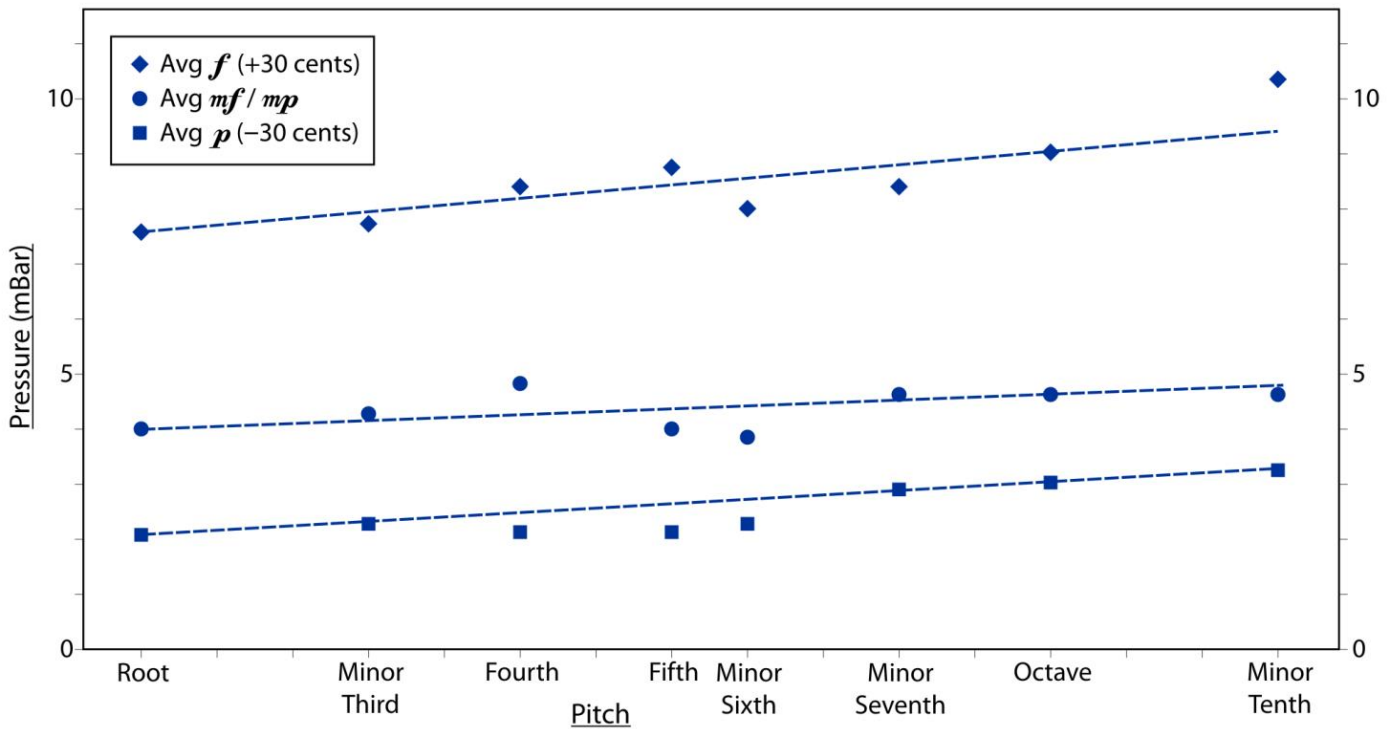

**Figure 10.** Breath pressure profile for a single Native American flute. *f* = *forte,* measured at reference pitch (RP) + 30 cents. *mf* / *mp* = *mezzo-forte / mezzo-piano,* measured at RP. *p* = *piano,* measured at RP – 30 cents. RP is concert pitch based on A=440.

pressure changes could not be measured. This may be particularly important for instruments such as the Slovakian fujara, where short air bursts are used to very briefly sound in the extreme upper registers of the instrument

However, the goal of this study was to evaluate various ethnic instrument classes against the potential health implications identified for Western classical instruments. It appears likely, at least for some of those health issues such as intermittent high-pressure glaucoma, that long-held tones are the primary agent.

Other limitations include:

- The use of a single subject, the investigator.

- The widely varying conditions across the set of prior studies surveyed for the purpose of comparing instrument characteristics.

- The subjective evaluation of tone in establishing a reference pitch in the case of instruments that were not tuned to concert pitch.

## Conclusions

This study was motivated by reports on a range of health issues associated with high intraoral pressure in some Western classical wind instruments and a lack of research in this area in ethnic wind instruments. Intraoral pressure was measured in six classes of ethnic wind instruments, and results were presented in the context of a survey of results across a broad range of wind instruments.

The results show that ethnic wind instruments, with the exception of ethnic reed instruments, have generally lower intraoral pressure requirements than Western classical wind instruments. The implication is that health issues that have been linked to high intraoral pressure in other studies are not an issue for these classes of ethnic instruments.

In the case of the Native American flute, the intraoral pressure requirements closely match the pressure involved in speech, a link that may be related to the instrument's roots in a tradition of poetic speech.

## References

For a general bibliography in the areas covered by this article, see http://www.Flutopedia.com/refs_bpress.htm.

## Appendix – Data Tables

The following pages provide numeric data for the charted data points that appear in the figures of this article.

**Table 1.** Musical Instrument Pressure Measurements Cited from the Literature

| Instrument | Source | N | Concert Pitch | Dyn | IOP (mBar) | Dyn | IOP (mBar) | Fig. |
|---|---|---|---|---|---|---|---|---|
| Alto Recorder | [Garcia 2010], Figure 5.13, graphic interpolation | 1 | $F_4$ | *ff* | 3.60 | *pp* | 1.25 | 5 |
| | | | $F_5$ | *ff* | 5.20 | *pp* | 2.60 | |
| | | | $F_6$ | *ff* | 17.75 | *pp* | 8.00 | |
| | | | $A_6$ | *ff* | 30.10 | *pp* | 9.40 | |
| Alto Saxophone | [Fuks 1996], Figures 16 and 18, graphic interpolation | 2 | $F_3$ | *ff* | 27.17 ± 2.80 | *pp* | 22.91 ± 0.77 | 5 |
| | | | $G_4$ | *ff* | 57.65 ± 22.91 | *pp* | 20.95 ± 2.87 | |
| | | | $A_5$ | *ff* | 38.94 ± 4.62 | *pp* | 19.05 ± 3.36 | |
| Bassoon | [Fuks 1996], Figures 28 and 30, graphic interpolation | 2 | $C_2$ | *ff* | 23.58 ± 1.04 | *pp* | 12.52 ± 0.94 | 4 |
| | | | $B_4$ | *ff* | 69.17 ± 19.09 | *pp* | 28.69 ± 5.11 | |
| Bombardon | [Stone-WH 1874] | 1 | $B_2$ | *mf* | 7.47[a] | | | 8 |
| | | | $G_4$ | *mf* | 89.67[a] | | | |
| Clarinet | [Fuks 1996], Figures 10 and 12, graphic interpolation | 2 | $F_3$ | *ff* | 45.97 ± 2.35 | *pp* | 28.61 ± 2.86 | 5 |
| | | | $D_5$ | *ff* | 51.95 ± 2.91 | *pp* | 26.00 ± 1.58 | |
| | | | $E_6$ | *ff* | 40.35 ± 4.49 | *pp* | 21.60 ± 1.18 | |
| Euphonium | [Stone-WH 1874] | 1 | $B_2$ | *mf* | 12.45[a] | | | 8 |
| | | | $G_4$ | *mf* | 67.25[a] | | | |
| French Horn | [Stone-WH 1874] | 1 | $G_3$ | *mf* | 7.47[a] | | | 8 |
| | | | $G_5$ | *mf* | 99.64[a] | | | |
| Oboe | [Fuks 1996], Figures 22 and 24, graphic interpolation | 2 | $C_4$ | *ff* | 51.96 ± 4.59 | *pp* | 37.97 ± 2.72 | 4 |
| | | | $E_6$ | *ff* | 109.13 ± 14.82 | *pp* | 53.21 ± 0.21 | |
| | [Adduci 2011], Table 14 | 4 | $D_4$ | *mf* | 41.32 ± 4.00 | | | 4 |
| | | | $G_4$ | *mf* | 43.80 ± 4.94 | | | |
| | | | $C_5$ | *mf* | 44.22 ± 4.52 | | | |
| | | | $A_5$ | *mf* | 52.85 ± 3.26 | | | |
| Trumpet | [Fletcher-NH 1999], Figures 1–3, logarithmic graphic interpolation | 3 | $C_2$ | *ff* | 51.96 ± 4.59 | *pp* | 37.97 ± 2.72 | 8 |
| | | | $B_4$ | *ff* | 109.13 ± 14.82 | *pp* | 53.21 ± 0.21 | |
| Western Concert Flute | [Fletcher-NH 1975], Figure 1, logarithmic graphic interpolation | 4 | $C_4$ | *ff* | 51.96 ± 4.59 | *pp* | 37.97 ± 2.72 | 5 |
| | | | $C_7$ | *ff* | 109.13 ± 14.82 | *pp* | 53.21 ± 0.21 | |
| | [Coltman 1966] and [Coltman 1968], as charted in [Fletcher-NH 1975], Figure 1, logarithmic graphic interpolation | 1 | $C_4$ | *mf* | 1.02 | | | not charted |
| | | | $C_7$ | *mf* | 9.16 | | | |
| | [Bouhuys 1965], as reported in [Fletcher-NH 1975], p. 233 | | $C_4$ | | 7.5–13 | | | not charted |
| | | | $A_6$ | | 27–40 | | | |
| | [Montgermont 2008], Figure 4d, graphic interpolation | 1 | $D_4$ | *ff* | 1.60 | *pp* | 0.79 | not charted |
| | | | $D_5$ | *ff* | 4.76 | *pp* | 2.58 | |
| | | | $D_6$ | *ff* | 9.15 | *pp* | 6.63 | |

*Note*: Intraoral pressure measurements from various sources were converted to millibars, in many cases, by physically interpolating the location of data points. In some cases, the graphs use a logarithmic scale, which needed a non-linear interpolation. Dyn = Dynamic. IOP = Intraoral pressure. Fig. = the number of the figure in this article on which the data for the row is charted. [a]Measurement given in whole inches $H_2O$.

**Table 2.** Vocal Pressure Measurements Cited from the Literature

| | Source | $N$ | Concert Pitch | Dyn | Pressure (mBar) | Dyn | Pressure (mBar) | Fig. |
|---|---|---|---|---|---|---|---|---|
| **Singing** | [Schutte 2003], Figure 3 – Average of 1974 and 1996 measurements, graphic interpolation | 1 | $A_2$ | *ff* | 7.12[s] | *pp* | 2.45[s] | 6 |
| | | | $E_3$ | *ff* | 12.73[s] | *pp* | 4.79[s] | |
| | | | $A_4$ | *ff* | 24.87[s] | *pp* | 7.00[s] | |
| | | | $E_4$ | *ff* | 50.20[s] | *pp* | 27.46[s] | |
| | | | $B^b_4$ | *ff* | 68.72[s] | *pp* | 52.82[s] | |
| | [Borch 2011], Figure 7, graphic interpolation | 1 | $F^{\#}_4$ | Rock music | 53.53[b] | | | 6 |
| | | | $G_4$ | Soul music | 29.92[b] | | | |
| | | | $D_4$ | Pop music | 24.37[b] | | | |
| **Speech** | [Subtelny 1966], Table 1 | 10 males | $G_2$–$G_4$[d] | *"natural and comfortable level"* | 6.64 ±0.99[e] | | | 3 |
| | | 10 females | $A_3$–$G_5$[d] | | 7.37 ±2.13[e] | | | |
| | | 10 children[c] | $C_4$–$A_5$[d] | | 10.28 ±3.57[e] | | | |
| | [Hodge-FS 2001], Figure 1c, graphic interpolation | | $G_2$–$G_4$[d] | *ff*[f] | 10.40 ±2.80 | | | 3 |
| | [Enflo 2009], Figure 5, graphic interpolation | 9 males | $F_2$ | *ppp* | 0.74[g] | | | 3 |
| | | | $F_3$ | *ppp* | 1.55[g] | | | |
| | | 6 females | $F_3$ | *ppp* | 0.72[g] | | | |
| | | | $A_4$ | *ppp* | 2.37[g] | | | |

*Note*: Pressure measurements are for intraoral pressure, unless indicated. Measurements were converted to millibars, in many cases, by physically interpolating the location of data points. Dyn = Dynamic. Fig. = the number of the figure in this article on which the data for the row is charted. [s]Subglottal pressure. [b]Highest subglottal pressure for the genre of music. [c]Age 6–10 years. [d]The range of pitches for typical speech is from [Williams-J 2010]. [e]Intraoral pressure speaking the phoneme that causes the highest intraoral pressure across 15 consonants: «tʃ» ("ch") in «itʃi» for males and children and «p» in «ipi» for females. [f]*"75% of dynamic range"*. [g]Measurement of the quietest possible speech.

**Table 3.** Native American Flute – Intraoral Pressure

| Concert Pitch | $n$ | Pressure (mBar) | | | |
|---|---|---|---|---|---|
| | | *f*-max | *f*-mean | *p*-mean | *p*-min |
| | | RP + 30 cents | | RP – 30 cents | |
| $E_3$ | 3 | 5.93 | | 1.65 | |
| $F\#_3$–$B_3$ | 21 | 6.69 | 4.94 | 1.47 | 0.83 |
| $C_4$–$F_4$ | 54 | 9.51 | 5.86 | 1.63 | 1.03 |
| $F\#_4$–$B_4$ | 144 | 14.13 | 6.63 | 2.28 | 0.90 |
| $C_5$–$F_5$ | 181 | 15.51 | 8.32 | 2.98 | 0.97 |
| $F\#_5$–$B_5$ | 144 | 20.55 | 9.66 | 3.71 | 1.45 |
| $C_6$–$F_6$ | 48 | 17.03 | 8.92 | 4.04 | 2.00 |
| $F\#_6$–$B_6$ | 10 | 11.79 | 9.93 | 5.26 | 3.93 |

*Note*: Intraoral pressure measurements across all Native American flutes ($N = 71$) are grouped by concert pitch into half-octave ranges. $n$ = the number of measurements for that pitch range. RP = reference pitch, concert pitch based on A=440 or, for instruments not tuned to concert pitch, a breath pressure that subjectively produced a good tone.

**Table 4.** Ethnic Duct Flutes – Intraoral Pressure

| Concert Pitch | $n$ | Pressure (mBar) | | |
|---|---|---|---|---|
| | | *f*-max RP + 30 cents | *mf*-mean RP | *p*-min RP – 30 cents |
| $F\#_3$–$B_3$ | 1 | 0.48 | 0.48 | 0.48 |
| $C_4$–$F_4$ | 14 | 3.03 | 1.98 ± 0.65 | 0.55 |
| $F\#_4$–$B_4$ | 34 | 6.00 | 2.64 ± 1.05 | 0.83 |
| $C_5$–$F_5$ | 73 | 23.86 | 3.86 ± 3.61 | 0.55 |
| $F\#_5$–$B_5$ | 76 | 25.51 | 5.05 ± 3.38 | 1.31 |
| $C_6$–$F_6$ | 53 | 30.34 | 7.83 ± 4.46 | 1.93 |
| $F\#_6$–$B_6$ | 36 | 43.02 | 13.54 ± 7.13 | 3.38 |
| $C_7$–$F_7$ | 19 | 36.27 | 18.16 ± 8.19 | 6.07 |
| $F\#_7$–$B_7$ | 2 | 47.23 | 45.44 ± 1.79 | 43.64 |

*Note*: Intraoral pressure measurements across all ethnic duct flutes ($N = 46$) are grouped by concert pitch into half-octave ranges. $n$ = the number of measurements for that pitch range. RP = reference pitch, concert pitch based on A=440 or, for instruments not tuned to concert pitch, a breath pressure that subjectively produced a good tone. Measurements were taken as with Native American flutes. Most of these flutes use the fingering 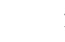 to produce the octave not in the second register. In addition, the fundamental note in each of the higher registers was attempted, as high as was possible on the instrument. These pressure measurements in the higher overtone registers were taken by establishing the pitch and then reducing breath pressure slightly to a point where the tone was stable — reference to precise tuning was not used in these higher overtone registers. The mean for each pitch range is taken over measurements at the reference pitch.

**Table 5.** Ethnic Reed Instruments – Intraoral Pressure

| Concert Pitch | $n$ | Pressure (mBar) | | |
|---|---|---|---|---|
| | | Maximum | Mean | Minimum |
| $C_4$–$F_4$ | 2 | 49.99 | 43.09 ± 6.89 | 36.20 |
| $F\#_4$–$B_4$ | 3 | 53.09 | 50.03 ± 2.87 | 46.19 |
| $C_5$–$F_5$ | 3 | 56.54 | 42.61 ± 11.26 | 28.96 |
| $F\#_5$–$B_5$ | 1 | 41.51 | 41.51 | 41.51 |

*Note*: Intraoral pressure measurements across all ethnic reed instruments ($N = 4$) are grouped by concert pitch into half-octave ranges. $n$ = the number of measurements for that pitch range. All measurements were taken on-pitch (*mf*).

**Table 6.** Ethnic Reedpipe Instruments – Intraoral Pressure

| Concert Pitch | n | Pressure (mBar) | | |
|---|---|---|---|---|
| | | Maximum | Mean | Minimum |
| $C_4$–$F_4$ | 12 | 26.75 | 18.66 ±5.45 | 8.14 |
| $F^{\#}_4$–$B_4$ | 12 | 35.85 | 17.14 ±4.13 | 10.48 |
| $C_5$–$F_5$ | 12 | 28.06 | 17.83 ±5.27 | 7.93 |
| $F^{\#}_5$–$B_5$ | 1 | 22.82 | 22.82 | 22.82 |

*Note*: Intraoral pressure measurements across all ethnic reedpipe instruments ($N = 12$) are grouped by concert pitch into half-octave ranges. $n =$ the number of measurements for that pitch range. Most measurements were taken at the reference pitch (RP), and the reported mean is taken over those measurements. RP is concert pitch based on A=440 or, for instruments not tuned to concert pitch, a breath pressure that subjectively produced a good tone. On one instrument, it was possible to take measurements at RP ± 30 cents and the maximum and minimum values reflect those additional measurements.

**Table 7.** Ethnic Overtone Whistles – Intraoral Pressure

| Concert Pitch | n | Pressure (mBar) | | |
|---|---|---|---|---|
| | | Maximum | Mean | Minimum |
| $G_1$ | 1 | 0.90 | 0.90 | 0.90 |
| $G_2$ | 1 | 1.65 | 1.65 | 1.65 |
| $D_3$ | 1 | 2.07 | 2.07 | 2.07 |
| $F^{\#}_3$–$B_3$ | 7 | 3.17 | 1.29 ± 1.00 | 0.28 |
| $C_4$–$F_4$ | 3 | 5.24 | 3.38 ± 1.79 | 0.97 |
| $F^{\#}_4$–$B_4$ | 9 | 12.13 | 4.01 ± 3.90 | 0.69 |
| $C_5$–$F_5$ | 6 | 3.72 | 3.03 ± 0.52 | 2.28 |
| $F^{\#}_5$–$B_5$ | 8 | 8.07 | 4.95 ± 1.69 | 2.14 |
| $C_6$–$F_6$ | 13 | 17.86 | 10.16 ± 3.97 | 4.21 |
| $F^{\#}_6$–$B_6$ | 13 | 33.78 | 19.98 ± 6.48 | 7.86 |
| $C_7$–$F_7$ | 5 | 64.26 | 36.87 ± 15.46 | 20.75 |
| $F^{\#}_7$–$B_7$ | 1 | 40.33 | 40.33 | 40.33 |

*Note*: Intraoral pressure measurements across all ethnic overtone whistles ($N = 8$) are grouped by concert pitch into half-octave ranges. $n =$ the number of measurements for that pitch range. One pressure measurement was taken for each note in each register by establishing the pitch and then reducing breath pressure slightly to a point where the tone was stable — reference to precise tuning was not used for this class of instruments.

**Table 8.** Ethnic Overtone Flutes – Intraoral Pressure

| Concert Pitch | n | Pressure (mBar) | | |
|---|---|---|---|---|
| | | Maximum | Mean | Minimum |
| $F^{\#}_2$–$B_2$ | 2 | 0.69 | 0.62 ±0.07 | 0.55 |
| $F^{\#}_3$–$B_3$ | 2 | 1.65 | 1.55 ±0.10 | 1.45 |
| $C_4$–$F_4$ | 2 | 3.24 | 3.10 ±0.14 | 2.96 |
| $F^{\#}_4$–$B_4$ | 5 | 9.93 | 6.41 ±2.59 | 2.48 |
| $C_5$–$F_5$ | 8 | 15.31 | 7.89 ±5.30 | 1.52 |
| $F^{\#}_5$–$B_5$ | 11 | 30.68 | 14.84 ±11.43 | 2.07 |
| $C_6$–$F_6$ | 8 | 8.96 | 5.07 ±2.27 | 3.10 |
| $F^{\#}_6$–$B_6$ | 5 | 17.51 | 8.74 ±4.73 | 4.62 |
| $C_7$–$F_7$ | 4 | 21.79 | 13.77 ±5.47 | 7.38 |
| $F^{\#}_7$–$B_7$ | 1 | 14.75 | 14.75 | 14.75 |

*Note*: Intraoral pressure measurements across all ethnic overtone flutes ($N = 8$) are grouped by concert pitch into half-octave ranges. $n =$ the number of measurements for that pitch range. Measurements were taken at reference pitch (RP) in the lowest note. RP is concert pitch based on A=440 or, for instruments not tuned to concert pitch, a breath pressure that subjectively produced a good tone. Measurements for the fifth note were was taken using the fingering 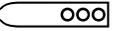 or 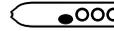 in the first register, regardless of what pitch was produced. The fundamental note in each of the higher registers was attempted, as high as was possible on the instrument. These pressure measurements in the higher overtone registers were taken by establishing the pitch and then reducing breath pressure slightly to a point where the tone was stable — reference to precise tuning was not used in these higher overtone registers.